\documentclass[prb,twocolumn,showpacs,bm,amssymb,amsmath,floatfix]{revtex4}
\usepackage{epsfig}
\begin{document}
\title{$F$-electron spectral function of the Falicov-Kimball model
in infinite dimensions: the half-filled case}
\date{\today}
\author{J. K. Freericks}
\email{freericks@physics.georgetown.edu}
\homepage{http://www.physics.georgetown.edu/~jkf}
\affiliation{Department of Physics, Georgetown University, Washington, DC
20057}
\author{V. M. Turkowski}
\email{turk@physics.georgetown.edu}
\homepage{http://www.physics.georgetown.edu/~turk}
\affiliation{Department of Physics, Georgetown University, Washington, DC
20057}
\author{V. Zlati\'c}
\email{zlatic@ifs.hr}
\homepage{http://bobi.ifs.hr/~zlatic}
\affiliation{Institute of Physics, Bijenicka c. 46, P. O. B. 304, 10000 
Zagreb, Croatia}

\begin{abstract}
The $f$-electron spectral function of the Falicov-Kimball model is 
calculated via a Keldysh-based many-body formalism
originally developed by Brandt and Urbanek.  We provide results for both
the Bethe lattice and the hypercubic lattice at half filling.  Since the 
numerical computations
are quite sensitive to the discretization along the Kadanoff-Baym contour and
to the maximum cutoff in time that is employed, we analyze the accuracy of the 
results using a variety of different moment sum-rules and spectral formulas.
We find that the $f$-electron spectral function has interesting temperature
dependence becoming a narrow single-peaked function for small $U$ and
developing a gap, with two broader peaks for large $U$.
\end{abstract}

\pacs{71.10.-W, 71.27.+a, 71.30.+h}
\maketitle

\section{Introduction}

Nonequilibrium many-body physics is becoming an increasingly important field
because it allows strongly correlated electrons to be examined in the
presence of large external fields, which can drive them far from equilibrium;
as nanotechnology research grows, there are an increasing number of nanoscale
solid-state devices which are exposed to extreme fields. The formalism to solve
nonequilibrium problems was developed independently by Kadanoff and 
Baym\cite{kadanoff_baym_1962} and by Keldysh\cite{keldysh_1964}. In principle,
it allows
these nonequilibrium problems to be solved exactly (including all nonlinear
field effects), but usually
the formalism is employed in a perturbative approach for the electron
correlations. These nonequilibrium formalisms can also be used to calculate
equilibrium Green's functions, especially in cases where alternative
analytic continuation methods are intractable.  We examine the simplest
such problem---the $f$-electron spectrum of the Falicov-Kimball model. This
problem can be solved exactly with dynamical mean field theory (DMFT).

The spinless Falicov-Kimball model\cite{falicov_kimball_1969} (FK) 
describes the dynamics of two types of electrons:  
conduction electrons (created or destroyed at site $i$ by 
$d^{\dagger}$ or $d$) 
and localized electrons (created or destroyed at site $i$ by 
$f^{\dagger}$ or $f$). 
The non-interacting conduction electrons can hop 
between nearest-neighbor sites on a D-dimensional lattice, 
with a hopping matrix  $-t_{ij}=-t^*/2\sqrt{D}$; 
we choose a scaling of the hopping matrix that yields a nontrivial limit
in infinite-dimensions\cite{metzner_vollhardt_1989}.
The f-electrons have a site energy $E_f$, 
and a chemical potential $\mu$ is employed to adjust the 
total number of electrons $n_d+n_f=n_{tot}$. 
The d- and f-number operators at each site are $n_d$ and $n_f$. 
There is a finite Coulomb interaction $U$ between d- and f-electrons
that occupy the same lattice site, and so the 
Falicov-Kimball Hamiltonian for the lattice is 
\begin{eqnarray}
{\mathcal H}_{FK}&=&
\sum_{ij }(-t_{ij}-\mu \delta_{ij})d_i^{\dagger}d_{j}
+
\sum_{i}(E_f-\mu )f_{i}^{\dagger }f_{i}\nonumber\\
&+&
\sum_iUd_i^{\dagger}d_if_i^{\dagger }f_i. 
                                                  \label{H_FK}
\end{eqnarray}
The FK lattice model [in Eq.~(\ref{H_FK})] can be solved in infinite 
dimensions using the methods of Brandt and Mielsch\cite{brandt_mielsch_1989}.
We consider two kinds of lattices: 
(i) the hypercubic lattice with a Gaussian noninteracting density of states 
$\rho_{hc}(\epsilon) =\exp [-\epsilon^2/t^{*2}]/(\sqrt{\pi}t^*)$,  
and (ii) the infinite-coordination Bethe lattice with a semicircular noninteracting density of states
$\rho_B(\epsilon)=\sqrt{4t^{*2}-\epsilon^2}/(2\pi t^{*2})$; we take $t^*$ as the unit of energy ($t^*=1$),
and consider only the homogeneous phase, where all quantities are translationally invariant. 

Despite the conservation of the local $f$-electron number
($[\mathcal{H}_{FK},f^\dagger_i,f_i]=0$), the $f$-electrons
have nontrivial dynamics as a function of $T$\cite{czycholl_1999,%
freericks_zlatic_RMP_2003}, so we 
expect the $f$-electron spectral function to have an interesting evolution 
with the correlation strength $U$.  When $U=0$, the spectral function is a delta
function, which broadens as $U$ increases because the $f$-electron sees a
fluctuating $d$-electron that hops onto and off of the local site as a function
of time. When $U$ is increased further, a metal-insulator transition takes
place, and we expect the $f$-spectral function to develop a gap as
$T\rightarrow 0$.  Hence the $f$-electron spectral function should have
rich behavior as a function of $U$ and $T$.

The organization of this paper is as follows. In Section II, we present the 
formalism, outlining in detail how the analytic continuation is carried out
within this nonequilibrium approach; our notation is nonstandard because we
use the usual Kadanoff-Baym contour, but we stick with the Green's function
definitions for $G^>$ and $G^<$ of Brandt and Urbanek\cite{brandt_urbanek_1992}.
In Section III, we present our numerical
results for both the Bethe and hypercubic lattices, and we conclude in
Section IV.  A short communication, showing the spectral function at 
the critical or near critical value of $U$ for the metal-insulator
transition has been completed\cite{freericks_turkowski_zlatic_2004a},
and a longer contribution, detailing the computational algorithm, 
parallelization, and numerical accuracy, has also been 
completed\cite{freericks_turkowski_zlatic_2004b}.

                         \section{Formalism}

The many body problem on an infinite-coordination lattice can be solved by a mean-field-like  
procedure, because the self energy of the conduction electrons is 
local\cite{metzner_vollhardt_1989}. Hence, the local 
d-electron Green's function $G_{loc}(z)$ on the lattice satisfies
\begin{equation}
G_d^{loc}(z) 
=\int \frac{\rho(\epsilon)}
{z+\mu-\Sigma_d(z)-\epsilon}d\epsilon,
                                            \label{eq: gloc}
\end{equation}
where $z$ is a complex variable, and $\Sigma_d$ is the 
momentum-independent self energy.
As noted by Brandt and Mielsch\cite{brandt_mielsch_1989}, the lattice 
self energy coincides with the self energy of an atomic d-state 
coupled to an f-state with the same Coulomb interaction as on the 
lattice,  and perturbed by an external time-dependent field, 
$\lambda(\tau,\tau^{\prime})$, which mimics the hopping on the lattice. 
For an appropriate choice of the $\lambda$-field, 
the functional dependence of $\Sigma_d$ on $G_d(z)$ 
and $G_f(z)$, the atomic propagators for d- and f-states, 
is exactly the same as in the lattice case. 
The lattice problem is thus reduced to finding the atomic 
self-energy functional for the d-electrons, and then setting 
$G_{d}^{loc}(z) = G_{d}(z)$ and $G_{f}^{loc}(z) = G_{f}(z)$ on each lattice site. 

The FK atom can be solved by using the interaction representation, 
such that the time dependence of operators is defined by the atomic 
Hamiltonian, 
\begin{equation}
                                         \label{H_atom}
{\mathcal H}_{\rm at}
=
 - \mu d^{\dagger }d
 +
 (E_f-\mu )f^{\dagger }f
 +
 U  d^{\dagger }d
 f^{\dagger }f, 
  \end{equation}
and the time-dependence of the state vectors is governed by an  
evolution operator which is defined by the $\lambda$-field. 
For purely imaginary times, $\bar \tau \in (0,-i\beta)$,  the evolution operator is 
\begin{equation}
                                               \label{S-matrix}
S(\bar \tau, \lambda)
=
T_{\bar \tau}
e^{+ \int_0^{\bar \tau}d\bar \tau^{\prime}
        \int_0^{\bar \tau} d\bar \tau^{\prime\prime}
      \lambda(\bar \tau^{\prime},\bar \tau^{\prime\prime})
       d^{\dagger}(\bar \tau^{\prime})d(\bar \tau^{\prime\prime})} .
\end{equation} 
where $T_{\bar \tau} [\cdots] $ orders all the operators within the bracket with 
respect to the position on the line $(0,-i\beta)$ of their time-argument in such 
a way that the time-arguments which are closer to zero appear further to the right 
and  there is an overall change of sign whenever the time ordering interchanges 
two fermionic operators. The time evolution of the operators between an initial time 
$\bar \tau^\prime$ and the final time $\bar \tau$ 
is determined by ${\mathcal H}_{at}$ as,  
\begin{equation}
                                             \label{op-evolution}
O(\bar \tau)
=
e^{ i (\bar \tau-\bar \tau^\prime)  {\mathcal H}_{at}} O(\bar \tau^\prime) e^{-i(\bar \tau-\bar \tau^\prime) {\mathcal H}_{at} }. 
\end{equation} 
The integration is along the imaginary-time axis ($\bar \tau$ and $\bar \tau$ are purely imaginary), 
i.e., we have made Wick's rotation $\bar \tau=-i\tau$ with respect to the real  variable used by  
Brandt and Mielsch\cite{brandt_mielsch_1989}.
 
The atomic Hamiltonian ${\mathcal H}_{\rm at}$ and the time-dependent field 
$\lambda(\bar \tau,\bar \tau^\prime) $ define the partition function of the FK atom,  
\begin{equation}
{\cal Z}_{at}(\lambda,\mu,\beta)=  {\rm Tr}_{df} T_{\bar \tau} [\exp\{-\beta {\mathcal H}_{\rm at} \} S(\beta,\lambda)]   , 
                                          \label{eq:Z_at}
\end{equation}
where the trace is taken over the atomic d- and f-states. Since the number of f-electrons 
is a conserved quantity, we can write 
\begin{equation}
                                                 \label{Z_Z0-product}
 {\cal Z}_{at}(\lambda)
=
 {\cal Z}_0(\lambda,\mu)
+ e^{-\beta (E_f-\mu)}
 {\cal Z}_0(\lambda,\mu-U).
\end{equation}
with $Z_0(\lambda,\mu)$ the partition function of a d-electron subject to the 
$\lambda$-field in the absence of f-electrons ($n_f=0$). That is, 
$Z_0(\lambda,\mu)={\rm Tr}_d T_{\bar \tau} [\exp\{-\beta {\mathcal H}_{0}\}S(\beta,\lambda)]$, 
where ${\mathcal H}_0=-\mu d^\dagger d$. 

The field $\lambda(\bar \tau,\bar \tau^{\prime})$ gives rise to fluctuations in the d-occupancy,  
which correspond in the equivalent lattice problem to the local d-fluctuations due to the 
d-electron hopping.  We choose $\lambda(\bar \tau,\bar \tau^{\prime})$ such that it satisfies 
the same (anti-periodic) boundary condition as the imaginary-time Green's function 
(see below) and expand it in a Fourier series along the imaginary axis 
(recall $\bar\tau$ is purely 
imaginary), 
\begin{equation}
\lambda( { \bar \tau}-{\bar \tau^{\prime}} )
=
T\sum_n e^{ \omega_n( {\bar \tau}-{\bar \tau^{\prime}} ) }\lambda_n. 
\label{eq:lambda_FT}
\end{equation}
Here, $\omega_n=\pi(2n+1)T$ is the Fermionic Matsubara frequency and we set $k_B=1$.  
The $\lambda$-field  in the complex $\omega$-plane, which is needed for the f-electron propagator, 
can be determined by an iterative procedure using the DMFT self-consistency condition 
for the d-electron's Green's function. 

The d-electron Green's function is defined as a functional derivative of the 
atomic partition function\cite{kadanoff_baym_1962}, 
$G_d(\bar \tau-\bar \tau^\prime) = 
{\delta \ln {\cal Z}_{\rm at}} / {\delta \lambda(\bar \tau^\prime,\bar \tau)}$, 
which gives, 
\begin{equation}
G_{d}(\bar \tau-\bar \tau^{\prime})
=- \frac{1}{{\cal Z}_{at}}
{\rm Tr}_{df} T_{\bar \tau}  \left[
 e^{-\beta {\mathcal H}_{at}} S(\beta,\lambda) d(\bar \tau)d^{\dag}(\bar \tau^{\prime})
 \right]  . 
                               \label{eq:D_int}
\end{equation}
$G_{f}(\bar \tau-\bar \tau^{\prime})$ is periodic  on the imaginary time axis 
with period $2i\beta$, and is antiperiodic $\mod(i\beta)$. It depends on the difference 
of the time-arguments, because we are in thermal equilibrium and the system is 
time-translation invariant;  it has a discontinuity at $\bar \tau=\bar \tau^{\prime}$ 
and is therefore a non-analytic  function of $\bar \tau-\bar \tau^{\prime}$.



  




To find $G_{f}$ we  define the effective-medium Green's function, 
$G_d^0(\bar \tau-\bar \tau^\prime) = 
{\delta \ln {\cal Z}_{0}} / {\delta \lambda(\bar \tau^\prime,\bar \tau)}$, which reads 
\begin{equation}
G_d^0(\bar \tau-\bar\tau^\prime)
=- \frac{1}{ {\cal Z}_{0}(\lambda,\mu)}
{\rm Tr}_{d} T_{\bar \tau}  \left[
 e^{-\beta {\mathcal H}_{0}} S(\beta,\lambda) d(\bar \tau)d^{\dag}(\bar \tau^{\prime})
 \right]  , 
                               \label{eq:D_eff_med}
\end{equation}
and satisfies the equation of motion [EOM]\cite{kadanoff_baym_1962}  
\begin{equation}
\int d\bar{\bar\tau} [(-\partial_{\bar\tau}+i\mu)\delta_c(\bar\tau- \bar{\bar\tau}) 
+
\lambda(\bar\tau-\bar{\bar\tau})]G_d^0(\bar{\bar\tau}-\bar\tau^\prime)
=\delta_c(\bar\tau-\bar\tau^\prime)
                                                                       \label{eq: eom_geff}
\end{equation}
with $\delta_c(\bar\tau)$ the delta function defined on the line segment
$(0,-i\beta)$ with the normalization $\int d\bar\tau
\delta_c(\bar\tau)=1$ (i.e., $\delta_c(\bar\tau)=iT\sum_n\exp[\omega_n\bar\tau]$).
Since $G_d^0$ also satisfies the usual periodic boundary conditions, Fourier
transforming the EOM gives the solution, 
\begin{equation}
                                        \label{G_0^(-1)_omega}
[G_{d}^{0}(i\omega_n)]^{-1}
={i\omega_n+\mu - \lambda_n},
\end{equation}
and the full Green's function follows 
as\cite{brandt_mielsch_1989,freericks_zlatic_RMP_2003} 
\begin{equation}
G_d(i\omega_{n})= 
\frac{w_0}{[G_{d}^{0}(i\omega_n)]^{-1}}
+ 
\frac{w_1}{[G_{d}^{0}(i\omega_n)]^{-1}-U},
                                             \label{G-atomic}
\end{equation}
with $w_0$ and $w_1$ the statistical weighting factors of the unoccupied and the 
occupied f-states, respectively. These are given by\cite{brandt_mielsch_1989} 
$w_1=1-w_0$ and $w_0 = {\cal Z}_0/{\cal Z}_{at}$, where 
\begin{equation}
                                        \label{solution_0}
 {\cal Z}_0(\lambda,\mu) 
 = 
2e^{\beta\mu/2}
  \prod_n \frac{[G_{d}^{0}(i\omega_n)]^{-1}}{i\omega_n}.
\end{equation}
Defining the self-energy function of the FK atom $\Sigma_{d}$ 
by Dyson's equation, 
\begin{equation}
\Sigma_{d}
=
[G_{d}^{0}]^{-1} 
-
[G_{d}]^{-1} , 
                                           \label{Dyson}
\end{equation}
which holds on the imaginary and the real frequency axes, 
we can find the $\lambda$-field and the statistical weighting factors $w_0$ and $w_1$ 
by an iterative procedure. We start with a trial self-energy on the imaginary axis and 
calculate $G_d$ from Eq.~(\ref{eq: gloc}), get $[G_{d}^{0}]^{-1} $ from Eq.~(\ref{Dyson}), 
calculate $w_0$ and $w_1$ using Eqs.~(\ref{solution_0})  and (\ref{Z_Z0-product}), 
recalculate $G_d$ from (\ref{G-atomic}), and find the new $\Sigma_{d}$ from Eq.~(\ref{Dyson}). 
Once the procedure is converged on the imaginary axis, we have the weights $w_0$ 
and $w_1$, the chemical potential $\mu$, and the self-consistent solution for the $\lambda$-field. 
Since  $w_0$ and $w_1$ are just numbers, we can analytically continue 
$G_d(i\omega_{n})$ given by Eq.~(\ref{G-atomic}) from the imaginary axis 
in the complex frequency-plane and repeat the iterative procedure to find retarded 
quantities and the spectral function of the $\lambda$-field. The knowledge of the $\lambda$-field 
everywhere in the complex $\omega$-plane is a necessary
input to find the Green's function of the localized f-electrons. 

The imaginary time Green's function of the f-electrons  is 
defined by an expression analogous to Eq.~(\ref{eq:D_int})
\begin{equation}
G_{f}(\bar \tau-\bar \tau^{\prime})
=- \frac{1}{{\cal Z}_{at}}
{\rm Tr}_{df} T_{\bar \tau}  \left[
 e^{-\beta {\mathcal H}_{at}} S(\beta,\lambda) f(\bar \tau)f^{\dag}(\bar \tau^{\prime})
 \right]  , 
                 \label{eq:F_int}
\end{equation} 
it satisfies the same boundary condition, and has the same 
analytic structure as $G_{d}(\bar \tau-\bar \tau^{\prime})$. 
As usual, we express $G_{f}(\bar \tau-\bar \tau^{\prime})$  in terms of two analytic 
functions of $\bar \tau-\bar \tau^{\prime}$, such that 
\begin{equation}
G_{f}(\bar \tau-\bar \tau^{\prime})=
\left\{ \begin{array}{ll}
              G_{f}^{>}(\bar \tau-\bar \tau^{\prime}) & \mbox{for Im$(\bar \tau-\bar \tau^{\prime})< 0$} \\
              G_{f}^{<}(\bar \tau-\bar \tau^{\prime}) & \mbox{for Im$(\bar \tau-\bar \tau^{\prime})> 0$}   
\end{array}\right. 
.
\label{eq:F><}
\end{equation}
To find  $G_{f}^{>}$ and $G_{f}^{<}$ we introduce the real variables $\tau=i\bar \tau$ and 
$\tau^\prime=i\bar \tau^\prime$, and represent $G_{f}(\bar \tau-\bar \tau^{\prime})$ 
by the Matsubara sum, 
\begin{equation}
 G_{f}(\bar \tau-\bar \tau^{\prime})=T\sum_n e^{ -i\omega_n (\tau-\tau^{\prime}) } G_f(i\omega_n)
\label{eq:FT}
 \end{equation}
For $\tau-\tau^{\prime} \in (-\beta,0)$, 
which corresponds to  Im$(\bar \tau-{\bar \tau}^{\prime}) \in (0,\beta)$, 
 we replace the sum over Matsubara frequencies by a contour integral in the complex 
 frequency-plane. Using Res$\left. f(z)\right|_{i\omega_n}=- 1/\beta$, 
where $f(\omega)$ is the Fermi function, we obtain 
\begin{equation}
 G_{f}^{<}(\bar \tau-\bar \tau^{\prime})=
 \int_{-\infty}^{\infty} 
 d\omega  f(\omega) 
 e^{ -\omega (\tau-\tau^{\prime}) } A(\omega), 
 \label{eq:G<FT}
  \end{equation}
where $A(\omega)=- {\rm Im} G_f^R(\omega)/\pi$ is the spectral function. 
Here, we used the fact that $G_f(z)$ coincides with the 
retarded Green's function, $G_f^R(z)$,  in the upper half-plane and 
with the advanced Green's function, $G_f^A(z)$,  in the lower half-plane, 
and that the discontinuity of $G_f(z)$ across the cut along the real $\omega$-axis  is given by 
$[G_f^R(\omega+i0^+) - G_f^A(\omega-i0^+)]/2\pi i= -{\rm Im} G_f^R(\omega)/\pi $.  
Since $\tau-\tau^{\prime} \in (-\beta,0)$, the integrand is well behaved for 
$\omega\longrightarrow\pm\infty$ (the cutoff at $\omega\longrightarrow\infty$ is 
provided by the Fermi  function) and the integral defines an analytic 
function (it has derivatives to all orders).  
Reinstating the imaginary times  $\bar \tau$ and $\bar \tau^{\prime}$, we obtain 
\begin{equation}
 G_{f}^{<}(\bar \tau-\bar \tau^{\prime})=
  \int_{-\infty}^{\infty} 
 d\omega  A(\omega)  f(\omega) e^{ -  i\omega (\bar \tau-\bar \tau^{\prime}) } , 
 \label{eq:Ft<} 
  \end{equation}
which can be used to perform the analytic continuation from the imaginary to real times, 
$\bar \tau \longrightarrow t$, $\bar \tau^{\prime} \longrightarrow t^{\prime}$. 
Defining the Fourier transform of the real-time Green's function as,  
 \begin{equation}
 G_{f}^{<}(t-t^{\prime})=  
 \frac{1}{2\pi} 
  \int_{-\infty}^{\infty} 
 d\omega  e^{ - i \omega (t-t^{\prime}) }  G_{f}^{<}(\omega)
  \label{eq:F<}
  \end{equation}
it follows that 
  \begin{equation}
{G_{f}^{<}(\omega)}= 2\pi A(\omega)  f(\omega) ; 
  \end{equation}
hence our definition for $G^<$ and $G^>$ is missing a factor of $i$ from the
standard definition, but agrees with that of Brandt and 
Urbanek\cite{brandt_urbanek_1992}.
Similarly, for  $\tau-\tau^{\prime} \in (0,\beta)$ and  
Im$(\bar \tau-{\bar \tau}^{\prime}) \in (-\beta,0)$, 
we use Res$\left. f(-z)\right|_{i\omega_n}=1/\beta$ and
express the Matsubara sum for $G_{f}^{>}(\tau-\tau^{\prime})$ as, 
\begin{equation}
 G_{f}^{>}(\bar \tau-\bar \tau^{\prime})
 = -
\int_{-\infty}^{\infty} 
 d\omega A(\omega) f(-\omega) e^{ -i\omega (\bar \tau-\bar \tau^{\prime}) } . 
  \label{eq:Ft>}   
  \end{equation}
This integrand is also well behaved and shows that $G_{f}^{>}$ is an analytic function 
of imaginary times. The analytic continuation to the real axis, 
$\bar \tau \longrightarrow t$, $\bar \tau^{\prime} \longrightarrow t^{\prime}$, gives
\begin{equation}
 G_{f}^{>}(t-t^{\prime})
 = \int_{-\infty}^{\infty}   d\omega  A(\omega)
 [f(\omega) -1] e^{ - i \omega (t-t^{\prime}) } 
                            \label{eq:F>>}
  \end{equation}
  so that  the Fourier transform of $ G_{f}^{>}(t)$ reads,  
\begin{equation}
 G_{f}^{>}(\omega)={2\pi }  A(\omega) { [f(\omega)-1] }
\label{eq:F>}
\end{equation}
At half-filling, where $A(\omega) =A(-\omega) $, we use $G_{f}^{>}(t-t^{\prime})= [G_{f}^{>}(t^{\prime}-t)]^*$ 
and obtain from the inverse of Eq.~(\ref{eq:F>>}) the result, 
\begin{equation}
  A(\omega) = - \frac{2}{\pi} 
\int_{0}^{\infty}  d t  \cos( \omega t ) 
 \textrm{Re}  G_{f}^{>}(t).
 \end{equation}
Thus, the time-ordered  Green's function at real times can be written as, 
\begin{equation}
 G_{f}^{}(t-t^{\prime})=\int d\omega A(\omega) [f(\omega) -\Theta(t-t^{\prime})] 
 e^{ - i \omega (t-t^{\prime}) }  , 
\end{equation}
with $\Theta(x)$ the unit step function $\Theta(x>0)=1$ and $\Theta(x<0)=0$.

However, these formal manipulations, which reveal the analytic properties of 
the Green's function, do not  explicitly provide the spectral function, $A(\omega)$, 
which is needed to find $G_{f}^{>}$ or $G_{f}^{<}$.  On the other hand, the imaginary-time 
formalism provides  numerical results for the Green's function at the  Matsubara 
frequencies\cite{brandt_mielsch_1989,zlatic_review_2001,freericks_zlatic_RMP_2003} but it
does not reveal the full analytic structure and it does not provide the spectral function. 
Thus, the real-time Green's function of the f-electron cannot be inferred directly from these formal 
and numerical results but has to be calculated separately. 

\begin{figure}[htb]
\epsfxsize=1.7in
\epsffile{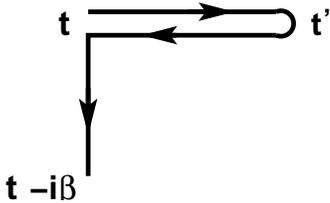}
\caption{\label{fig: contour}Kadanoff-Baym contour for evaluating the equilibrium Green's function for $t \leq t^{\prime}$. For $t \geq t^{\prime}$ the contour starts at  $t^\prime$ and runs to  $t$,  
then goes back to $t^\prime$ and ends at $t^\prime-i\beta$.}
\end{figure}

To obtain the real-time properties of the f-electron we define the contour-ordered 
Green's function in the interaction representation as 
\begin{equation}
G_{f}^c(t-t^{\prime})
=- \frac{1}{{\cal Z}_{at}}
{\rm Tr}_{df} 
T_c  \left[
 e^{-\beta {\mathcal H}_{at}} S_c(\lambda_c)
 f(t)
 f^{\dag}(t^{\prime})
 \right] , 
                        \label{eq:F_real}
\end{equation}
where 
\begin{equation}
                               \label{eq:S-labda_c}
S_c(\lambda_{c})
=
T_{c}
e^{\int_{ c} d\bar t \int_{c} d{\bar t}^{\prime}
       \lambda_{c}(\bar t,\bar t^{\prime})
       d^{\dagger}(\bar t)d(\bar t^{\prime})} ,
\end{equation}
is the analytic continuation of the evolution operator 
in Eq.~(\ref{S-matrix}) from imaginary times to  
times  on the contour which is depicted in Fig.~\ref{fig: contour} for the case 
$t^\prime\geq t$.  
For $t < t^\prime$  the contour starts at $t$ runs to $t^\prime$, goes back to $t $, and ends at $t-i\beta$. 
Once again, this notation is missing a factor of $i$ in the exponent of the 
evolution operator and as a prefactor for the Green's function, from that used
in standard approaches, but it agrees with Brandt and Urbanek\cite{brandt_urbanek_1992}.

%
%


The time dependence of the operators on the contour is defined as  
(assuming $ \bar t^\prime$ precedes $\bar t$ on $C$),  
\begin{equation}
                   \label{op-evolution-t}
O(\bar t)
=
e^{i(\bar t -\bar t^\prime){\mathcal H}_{at}} O(\bar t^\prime )e^{-i(\bar t  -\bar t^\prime){\mathcal H}_{at}}. 
\end{equation}
The $T_{c} \left[ \cdots \right]$ orders all operators within the 
bracket with respect to the position on the contour of their time arguments, 
such that, 
\begin{equation}
T_{ c} \left[  f(\bar t)  f^{\dag}( \bar t^{\prime}) \right]
=
\left\{ \begin{array}{ll}
               f( \bar t)  f^{\dag}( \bar t^{\prime})  & \mbox{ $ \bar t^{\prime}$ preceding $ \bar t$ on $ C$}  \\
             - f^{\dag}( \bar t^{\prime})  f( \bar t) & \mbox{ $ \bar t$ preceding $ \bar t^{\prime}$ on $ C$} 
\end{array}\right. 
,
\end{equation}
and similarly for the $d$ operators. 

The $\lambda$-field on the contour is obtained by an analytic continuation 
from the imaginary time axis.  Since the $\lambda$-field, considered as a function 
of imaginary times, satisfies the same boundary condition as the Green's function, 
it is composed of two analytic pieces which can be continued from imaginary times 
to complex times on the contour $C$. Starting from the Fourier transform 
given by Eq.~(\ref{eq:lambda_FT}) we obtain, in analogy with Eqs.~(\ref{eq:Ft<}) and (\ref{eq:Ft>}), 
the results, 
\begin{equation}
\lambda^<( { \bar \tau}-{\bar \tau^{\prime}} )
 = -\frac{1}{\pi} 
\int_{-\infty}^{\infty} 
 d\omega  f(\omega) e^{ -i\omega (\bar \tau-\bar \tau^{\prime}) }  {\rm Im} \lambda^R(\omega). 
  \end{equation}
 and 
\begin{equation}
\lambda^>( { \bar \tau}-{\bar \tau^{\prime}} )
 = -\frac{1}{\pi} 
\int_{-\infty}^{\infty}  d\omega  
[f(\omega) -1] e^{ -i\omega (\bar \tau-\bar \tau^{\prime}) }  {\rm Im} \lambda^R(\omega). 
  \end{equation}
where $ {\rm Im} \lambda^R(\omega)$ is the spectral function of the $\lambda$-field. 
These integrals define analytic functions of imaginary times and can be continued 
to the contour  shown in Fig.~\ref{fig: contour} simply by substituting $\bar \tau \rightarrow \bar t$
and $\bar \tau^\prime \rightarrow \bar t^\prime$, where $\bar t$ and $\bar t^\prime$ 
are on the contour $C$. 
Thus, the contour-ordered $\lambda$-field can be written as, 
\begin{eqnarray}
\lambda_c(\bar t, \bar t^\prime)
&=&
-\frac{1}{\pi}\int_{-\infty}^{\infty}d\omega
\textrm{Im} \lambda^R(\omega)\exp [-i\omega(\bar t-\bar t^\prime)]\cr
&\times&[f(\omega)- \theta_c(\bar t- \bar t^\prime)], 
\label{eq: lambda_t}
\end{eqnarray}
where $\Theta_c=0$ if $\bar t$ precedes $\bar t^{\prime}$ on $C$ and $\Theta_c=1$ otherwise, and $\bar t$ and $\bar t^\prime$ lie anywhere on the contour. 
Restricting $\bar t$ and $ \bar t^\prime$ to the vertical part of the contour, 
and using the anti-periodicity along the imaginary axis, 
we can make the inverse Fourier transform 
\begin{eqnarray}
\lambda(i\omega_n)=\lambda_n
&=&
i\int_0^{-i\beta} d\bar\tau 
\lambda_c(\bar \tau,0)e^{-\omega_n\bar\tau}\nonumber\\
&=&
i\int_0^{-i\beta}  d\bar\tau 
\lambda_c(\bar \tau,-i\beta)e^{-\omega_n\bar\tau}
                                     \label{eq: ft_of_lambda}
\end{eqnarray}
and recover the spectral formula
\begin{equation}
\lambda(i\omega_n) = 
-\frac{1}{\pi} 
\int_{-\infty}^{\infty}  d\omega  \frac {\mbox{Im } \lambda^R(\omega)}{i\omega_n -\omega}  . 
  \end{equation}
Contrary to the arguments of $G_{f}(t-t^{\prime})$, which define the boundaries of the 
horizontal piece of the contour, the arguments $\bar t$ and $\bar t^{\prime}$ of the 
contour-ordered $\lambda$-field can be anywhere on the contour $C$, 
\textit{so that the dynamical mean field
connects the real and the imaginary parts of the contour}. 
In  DMFT,  the $\lambda$-field originates from the electron hopping on the lattice 
and it is responsible not only for the propagation of particles in real times but for 
the thermalization of the system as well. Contrary to most non-equilibrium problems, 
in which the real field is switched on at some time $t_0$ and the integration along the vertical 
part of the Kadanoff-Baym contour can be neglected as one approaches
the steady state, in DMFT problems it is essential 
to integrate over the whole contour, because the hopping on the lattice (which gives rise 
to the $\lambda$-field) is always present. 


We can now  find the contour-ordered Green's function using the Kadanoff-Baym 
EOM methods, and we consider the case $t >  t^\prime$. 
The fermionic operators on the contour satisfy the EOMs, 
\begin{equation}
                                                                         \label{d_d_tau}
i\frac{ d }{d \bar t} d^\dag(\bar t) 
= 
[\mu - U  f^\dag(\bar t) f(\bar t)] d^\dag(\bar t),
\end{equation}
and 
\begin{equation}
                                                                     \label{d_f_tau}
i\frac{d }{d  \bar t} f( \bar t)
= 
[(E_f-\mu) + U d^\dag( \bar t) d( \bar t)]  f( \bar t).
\end{equation}
Since  $ f^\dag(\bar t) f(\bar t)=f^\dagger(0)f(0)$ does not change with 
time (i.e., it commutes with the Hamiltonian), 
Eq.~(\ref{d_d_tau}) has a simple solution, 
\begin{equation}
                                                    \label{d_tau}
d^\dag(\bar t) 
= 
e^{ i(\mu-U)(\bar t -  t^{\prime})  f^\dag( t^{\prime} ) f( t^{\prime} ) }
d^\dag( t^{\prime})  , 
\end{equation}
where $t^\prime$ is the initial time on the contour. 
The evolution of f-electrons is more complicated, because  the number of d-electrons 
fluctuates in time and  we can only write the solution as a contour-ordered product,  
\begin{equation}
                                                    \label{fff_tau}
f(\bar t) 
=
e^{-i(E_f-\mu)(\bar t - t^\prime)} 
S_c^\prime(\chi_{\bar t}) f( t^\prime), 
\end{equation}
where 
\begin{equation}
                                                    \label{S'}
S_c^\prime(\chi_{\bar t})  
= 
T_c
\exp \left\{ 
         \int_c  d \bar{\bar t}   \int_c d \bar{\bar t}^{\prime } 
                \chi_{\bar t} (\bar{\bar t},\bar{\bar t}^{\prime})
                  d^\dag (\bar{\bar t}) d(\bar{\bar t}^\prime) 
     \right\}  ,
\end{equation}
and 
\begin{equation}
                                               \label{chi_tau}
\chi_{\bar t} ( \bar{\bar t},\bar{\bar t}^{\prime})
=-iU \Theta_c (\bar t - \bar{\bar t})
\delta_c(\bar{\bar t}-\bar{\bar t}^{\prime}).
\end{equation} 
Thus,  we obtain 
\begin{eqnarray}
G_f^>( t -  t^\prime)
&=&-\frac {e^{-i(E_f-\mu)( t -  t^\prime)} } {{\cal Z}_{at}}\nonumber\\
&\times&
 {\rm Tr}_{d\,f}
\left[
e^{ -\beta {\mathcal H}_{at} } 
S_c(\tilde \lambda_c)
f( t^\prime) f^\dag( t^\prime)
\right] , 
                                           \label{SS'}
\end{eqnarray}
where $S_c(\tilde \lambda_c)$ is the evolution operator in the 
presence of the modified time-dependent potential which is due to the 
fluctuation in the number of d-electrons during the propagation 
of an f-electron  from the initial time $t^\prime$ to the final time $t$, 
\begin{equation}
                                             \label{chi_vau}
\tilde \lambda_c( \bar{\bar t}, \bar{\bar t}^{\prime})
=
\lambda_c( \bar{\bar t}, \bar{\bar t}^{\prime})
+
\chi_{ t} ( \bar{\bar t}, \bar{\bar t}^{\prime}). 
\end{equation} 
Note, the operator sequence $f( t^\prime) f^\dag( t^\prime)$ commutes with $S_c(\tilde \lambda_c)$ 
and  removes all the occupied f-states from the trace, so that the f-propagator  can be expressed in terms 
of a partition function of an effective d-electron,  
\begin{equation}
G_f^>( t - t^\prime)
                                                    \label{FS}
=-\frac {e^{-i(E_f-\mu)( t -  t^\prime)} } {{\cal Z}_{at}}
{\cal Z}_{}(\mu,\tilde\lambda_c) , 
 \end{equation}
where ${\mathcal H}_{0}$ defines 
the dynamics of a d-electron when there are no f-electrons and 
\begin{equation}
{\cal Z}_{}(\tilde\lambda_c)
={\rm Tr}_{d} \left[
 e^{-\beta {\mathcal H}_{0} } S_c(\tilde\lambda_c)
 \right]
 \end{equation}
 is the partition function of such an electron subject to the effective $\tilde \lambda_c$-field.
Because of time-translation invariance, we set $t^\prime=0$ from now on.

To find the partition function of a d-electron driven by the time-dependent  
$\tilde \lambda_c$-field  we use again functional derivative techniques to  
define an auxiliary  Green's function,  
\begin{equation}
                                 \label{g-functional_derivative}
 g^c(\bar t,\bar t^{\prime}) 
=
\frac{1}{{\cal Z}_{}(\tilde\lambda)}
\frac{\delta {\cal Z}_{}(\tilde \lambda_c)}
     {\delta \tilde\lambda(\bar t^\prime,\bar t)}, 
\end{equation}
such that, 
\begin{equation}
                                                          \label{g-definition}
g^c(\bar t,\bar t^{\prime})
= -\frac{1}{ {\cal Z} ({\tilde\lambda})}
{\rm Tr}_d \left[
T_c
e^{-\beta H_0}
d(\bar t) d^\dagger(\bar t^{\prime})
S_c({\tilde\lambda_c})
\right].
\end{equation}
Note, the operator dynamics on the contour is now defined by ${\mathcal H}_0$, 
\begin{equation}
                   \label{op-evolution-H_0}
O(\bar t)
=
e^{i(\bar t -\bar t^\prime){\mathcal H}_0} O(\bar t^\prime )e^{-i(\bar t  -\bar t^\prime){\mathcal H}_0}. 
\end{equation} 
Next,  we introduce an auxiliary contour-ordered Green's function 
for a d-electron driven by the $\chi_{ t}$-field in the absence of the 
$\lambda_c$-field, 
\begin{equation}
                                          \label{g_0-functional_derivative}
g^c_0(\bar t,\bar t^{\prime}) 
=
 \frac{ \delta \ln {\cal Z}_0(\chi_t) }
       {\delta {\chi_{ t}}(\bar t^{\prime},\bar t)}
\end{equation}
where 
\begin{equation}
                                                   \label{Z_0-chi-definition}
{\cal Z}_0(\chi_{ t})
=
{\rm Tr}_d \left[T_c 
e^{-\beta H_0} S_c(\chi_{ t})
\right]=1+e^{\beta\mu-iU t},
\end{equation} 
is the effective partition function of such a system. Functional differentiation gives, 
\begin{equation}
                                            \label{g_0-definition}
g^c_{0}(\bar t,\bar t^{\prime})
= -\frac{1}{ {\cal Z}_0 (\chi_{ t}) }
{\rm Tr}_d \left[
T_c
e^{-\beta H_0^\sigma}
d_\sigma(\bar t) d^\dagger(\bar t^{\prime})
S_c(\chi_{ t})
\right].
\end{equation}
The evaluation of ${\cal Z}_0(\chi_{ t})$ and $g^c_{0}$ is straightforward (for details 
see Refs.~\onlinecite{zlatic_review_2001} and \onlinecite{freericks_zlatic_RMP_2003}), 
because the evolution operator $S_c({\chi_{ t}})$ does not change the number of d-electrons 
and the Hilbert space for the d-states comprises only two states ($n_d=0$ and $n_d=1$). 
The Green's functions $g$ and $g_{0}$ depend explicitly on the contour-times 
$\bar t$ and $\bar t^{\prime}$, and implicitly on the external time $ t$
(recall we set $t^\prime=0$). 

Taking the time derivatives of $g^c(\bar t,\bar t^{\prime})$ and $g^c_0(\bar t,\bar t^{\prime})$ 
with respect to $\bar t$ we find, using Eq.~(\ref{op-evolution-H_0}),  the EOMs, 
\begin{equation}
                                                                 \label{EOM-integral}
\int_c d \bar t^{\prime}
[g^c]^{-1}(\bar t,\bar t^{\prime})
g^c(\bar t^{\prime},\bar t)
=  \delta_c(\bar t-\bar t^\prime), 
\end{equation}
where 
\begin{equation}
                                                     \label{g_0^(-1)_tau}
[g^c]^{-1}(\bar t,\bar t^{\prime})
=
[g^c_0]^{-1}(\bar t,\bar t^{\prime})
+
\lambda_c(\bar t,\bar t^{\prime}),
\end{equation}
and 
\begin{equation}
                                                 \label{g_00^(-1)_tau}
[g^c_0]^{-1}(\bar t,\bar t^{\prime})
=
(-\frac{\partial }{\partial\bar t} + i\mu)
\delta_c(\bar t-\bar t^\prime)
+
\chi_{ t} (\bar t,\bar t^{\prime}). 
\end{equation}
In operator form, this can be written as,  
\begin{equation}
                                        \label{integral-operator}
[g^c]^{-1}g^c= \mathbf{1},
\end{equation}
where the unit operator $\mathbf{1}$ has the matrix elements 
$\delta_c(\bar t - \bar t^\prime)$. 
The Dyson equation for the integral operator $g^c$ can, thus, be written as,  
\begin{equation}
[g^c]^{-1} = [g^c_0]^{-1}[1+ g^c_0 \lambda_c ] , 
\end{equation}
with the operator product implying an integration over the contour $C$.  From 
the definition of $g_c$ in terms of the functional derivatives of the partition function 
$ {\cal Z}(\tilde\lambda_c)$, it follows that, 
\begin{equation}
 {\cal Z}(\tilde\lambda_c)
= e^{{\rm Tr}~\ln {\, [g^c]^{-1}}}, 
\end{equation}
where the continuous trace of a contour-ordered operator is given by 
the line integral over the contour $C$. Using Dyson's equation, this can be written as, 
\begin{equation}
 {\cal Z}(\tilde\lambda_c)
={\cal Z}_0(\chi_{\bar t}) e^{ {\rm Tr}~\ln {\, (1+g^c_0 \lambda_c)} }.
\end{equation}
To approximate the continuous trace by a discrete one\cite{brandt_urbanek_1992}  
we expand the logarithm,  
\begin{equation}
 \ln {\, (1+g^c_0 \lambda_c)}  = \sum_n \frac{1}{n}(g^c_0 \lambda_c)^n, 
\label{eq: expansion}
 \end{equation}
and replace each contour integral by a discrete sum, using a discrete
quadrature rule
\begin{equation}
\int_c dt I(t)=\sum_{i=1}^N W_i I(t_i)
\label{eq: quadrature}
\end{equation}
with weights $W_i$ for the discrete set of times $\{t_1,\dots,t_N  \}$ on the contour $C$.
Then, the multiple integrals in Eq.~\ref{eq: expansion}
reduce to matrix multiplication, and we can use the usual expression from
linear algebra,
\begin{equation}
 {\rm Tr}~\ln {\, (1+g^c_0 \lambda_c)}  = \ln \textrm{Det} {\, (1+g^c_0 \lambda_c)} 
 \end{equation}
 where $\textrm{Det}$ represents an $N\times N$ determinant, we obtain  
 the final result, 
\begin{eqnarray}
G_f^>(t)&=&-\frac{1}{{\mathcal{Z}}_{}}e^{-i(E_{f}-\mu)t}
{\mathcal{Z}}_{0}(\chi_{\bar t})\\
&\times&
\textrm{Det} [W_i\{\frac{\delta_{ij}}{\Delta t_c} 
+ 
\sum_k g^c (t_i,t_k)W_k\lambda_c(t_k,t_j)\}] \nonumber 
                                 \label{eq: f_greater_t}
\end{eqnarray}
which we calculate numerically 
($1/\Delta t_c$ is the approximation to the delta function
on the contour $C$ with $\Delta t_c$ the width of the interval that includes the delta function;
for a [midpoint] rectangular quadrature rule, one takes $W_i=\Delta t_c$).

\section{Computational Results}

The numerical evaluation of the $f$-electron Green's function appears to
be a rather straightforward procedure:  one decides on a step size for the
real-time axis $\Delta t_{\textrm{real}}$ and for the imaginary-time axis
$\Delta t_{\textrm{imag}}$ of the Kadanoff-Baym contour
and then calculates out to the largest
time that is feasible within the limitations of the computational resources.
In the results presented here, we take $\Delta t_{\textrm{real}}$ to range 
from 0.1 to 0.0125.  We fix $\Delta t_{\textrm{imag}}=0.05$.  The cutoff
in time is always taken to be no larger than 80.  
In order to calculate the $f$-electron Green's function, we need to
calculate the determinant of a discretized matrix operator.
This is done by first diagonalizing the 
matrix, and then taking the product of all of the eigenvalues. This step
is the most time-consuming step of the calculation, because the matrix is
a general complex matrix, with no special symmetries, and the eigenvalues are
usually complex-valued (the maximal matrix size that we consider is 
about $2100\times 2100$). Since each time $t$ chosen to evaluate the Green's
function requires a new contour, 
the grid of points on the time axis, where $G_f^>(t)$
is generated, need not use the same spacing as the discretizing grid of each
Kadanoff-Baym contour used for
discretizing the continuous matrix operator. Usually, we use a time-grid
spacing of
0.2 or 0.1, because the Green's function does not normally have oscillations
that are on a finer scale than that on the time axis.
Once the Green's function has been calculated on
the time-axis grid, we perform a Fourier transform to calculate it on the 
real frequency
axis.  We first spline our real-time data (using a shape-preserving
Akuba spline) onto a real-time grid that is twenty times smaller than
the originally chosen time grid spacing.  Next, we numerically sum the (cosine)
Fourier transform of the real part of the Green's function to determine the
spectral function (which is possible only at half filling; for other
fillings the analysis is more complicated\cite{freericks_zlatic_RMP_2003}). 
More details of the numerics can
be found elsewhere\cite{freericks_turkowski_zlatic_2004b}.

The spectral function satisfies a number of important properties.  Since our
calculation is an equilibrium calculation (even though we are employing a
nonequilibrium formalism), the spectral function is nonnegative and
the integral of the spectral function over all frequency is equal to one.
Furthermore, the Green's function on the real-time axis approaches 
$w_1-1$ as $t\rightarrow 0$ and has an exponentially decaying (and
possibly oscillating) behavior at large times.  It also increases
quadratically in $t$ for small times with a curvature that is independent
of temperature.  Unlike the conduction-electron
DOS, which is independent\cite{vandongen_1992} of temperature, the $f$-electron
DOS evolves\cite{brandt_urbanek_1992}
with $T$. But, because the value at $t=0$ [and the first and second derivative 
of ${\rm Re}G^>_f(t)$] is the
same for all temperatures, we find that the deviations of the real-time Green's
functions (due to changes in the temperature)
increase at  large times.

When the DOS develops a gap at low temperature, the long-time behavior
of the Green's function develops significant oscillations, with 
an amplitude that can decay to zero very slowly.  This creates numerical
problems, since it implies that the cutoff in time needs to be large in order 
to be able to accurately determine the DOS.  Indeed, we will find that
this cutoff dependence limits our ability to accurately determine the
DOS at low temperature.

\begin{figure}[htb]
\epsfxsize=3.0in
\epsffile{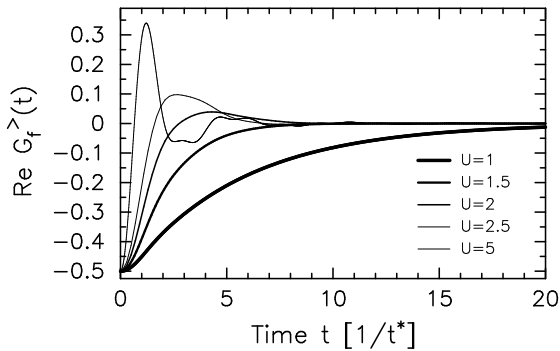}
\caption{\label{fig: G_time} Real part of the (greater)
$f$-electron Green's function
as a function of time for $T=1$ and five different values of $U$ on the
Bethe lattice.
}
\end{figure}

We begin our discussion with a plot of the real part of the (greater)
Green's function
versus time for $T=1$ on the Bethe lattice and for 5 different values of
$U$ (Fig.~\ref{fig: G_time}). 
The values of $\Delta t_{\textrm{real}}$ are 0.0125 for $U=1$, 0.025
for $U=2$ and $U=5$ and 0.05 for $U=2$ and $U=2.5$.  Notice how the
Green's function appears to have just a smooth exponential decay for small $U$,
but as $U$ increases, we first see the Green's function assume positive values,
and then we see that it picks up significant oscillations, whose period
decreases as $U$ increases.  When $U$ is small enough, that there is
no gap in the DOS, then we find that the long-time behavior is
exponentially decaying (with oscillations entering as the critical value
of $U$ for the Mott transition is approached). In this regime, we
can extrapolate the results for small time out to large time, by fitting the
Green's function tail with an exponential function, and evaluating that function
out to long times.  This allows us to use a smaller time cutoff (and thereby a 
smaller $\Delta t_{\textrm{real}}$), which
becomes increasingly important at low temperature in order to maintain
high quality in the data.

\begin{figure}[htb]
\epsfxsize=3.0in
\epsffile{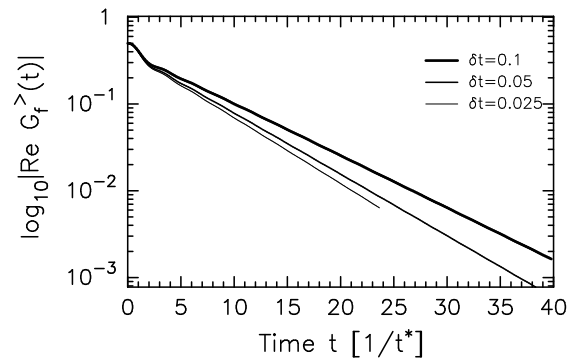}
\caption{\label{fig: log_G_time} Logarithm of the absolute value of
the real part of the (greater)
$f$-electron Green's function
as a function of time for $T=0.2$ and $U=1.5$ on the Bethe lattice.
Three different values of $\Delta t_{\textrm{real}}$ are shown.
}
\end{figure}

In Fig.~\ref{fig: log_G_time}, we plot the logarithm of the absolute value of
the real part
of the (greater) Green's function for $T=0.2$ and $U=1.5$ on the Bethe lattice.
Note how the tails of the Green's functions show perfect exponential 
decay at large time.  The exponent depends strongly on the discretization
$\Delta t_{\textrm{real}}$ of the Kadanoff-Baym
contour.  But because of this simple
exponential behavior, we can append extrapolated tails to our more accurate 
calculations (with smaller $\Delta t_{\textrm{real}}$) and construct
good Fourier transforms.  Unfortunately, there does not seem to be
any simple way to extrapolate the results to the limit $\Delta t_{\textrm{real}}
\rightarrow 0$ on the time axis.  The problem is that the systematic error
due to a finite $\Delta t_{\textrm{real}}$ is not arising from a Trotter
break-up which has a simple error term, but rather is arising from the
discretization of the continuous matrix operator.  When $U$ is large
enough that the tails of the Green's functions show oscillatory behavior
on top of the decaying behavior, we have no simple way to extrapolate
the tails out to large time, and the errors of the calculations become
larger. [One might have wanted to extrapolate $\ln |G_f^>(t)|$ on the time
axis to $\Delta t \rightarrow 0$ but that cannot be done once $G_f^>(t)$
crosses zero.]

When we perform the cosine Fourier transform of the real part of $G_f^>(t)$
to get the $f$-electron spectral function, we first spline the real time
data (with an exponentially decaying tail appended if appropriate)
onto a grid that is 20 times finer using a shape-preserving Akuba
spline, and then numerically perform the Fourier transform. 
Next we try to extrapolate the Fourier transform to
$\Delta t_{\textrm{real}}\rightarrow 0$. This extrapolation is performed using
an $n$-point Lagrange interpolation formula (pointwise
in $\omega$) on the different
DOS generated for the different $\Delta t_{\textrm{real}}$. Such a procedure
allows for higher order polynomial approximations to the extrapolation as
more data is included.  We sometimes find that even though we have data
for a number of different $\Delta t_{\textrm{real}}$ values, it is most
accurate to perform a linear extrapolation for the two smallest
values of $\Delta t_{\textrm{real}}$. We call this extrapolation
scheme $\delta$-extrapolation.

We judge the accuracy of our calculations in a number of different ways.
The first thing we do is to compare the moments of the DOS [with or without
an extra factor of $f(\omega)$] to exact results for those moments (see below).
We also compare the Matsubara frequency Green's functions (generated from
an independent program that works directly on the imaginary 
axis\cite{brandt_urbanek_1992,zlatic_review_2001,freericks_zlatic_RMP_2003}) with the Matsubara
frequency Green's function generated by integrating the spectral formula
(with the given DOS) for each Matsubara frequency
\begin{equation}
G_f(i\omega_n)=\int d\omega A_f(\omega)\frac{1}{i\omega_n-\omega}.
\label{eq: mats_spectral}
\end{equation}

One of the important checks of our numerical accuracy comes from a careful
comparison of the calculated results with a number of different moment
sum-rules of the DOS.  The sum rules can be derived in a straightforward
fashion: the DOS is first expressed as the imaginary part of the
Fourier transform of the real-time retarded Green's function.  By introducing
complete sets of states, the time dependence of the operators can be
expressed in terms of the many-body energies of the different states
(note that because this calculation is performed for the impurity, one
must include the evolution operator of the $\lambda$ field, but since it
commutes with the $f$-electron operators, it provides no further complications).
These can then be integrated over time, and when the imaginary part is taken, 
one gets a delta function in frequency, which allows the frequency integral
to be performed.  Finally, any energy factors that remain can be replaced by the
Hamiltonian, and the sums over the complete sets of states can be performed.
In the end, we are left with operator averages to evaluate.  These
results are summarized in Table~\ref{table: sum-rules}.  Note that when we 
perform actual calculations, we always add a small shift to the DOS in order to
satisfy the zero moment sum rule to at least one part in $10^5$ (typically this
means adding a shift no larger than 0.005 to the spectral function).

\begin{table*}
\caption{\label{table: sum-rules}
Sum rules for the $f$-electron DOS. The expectation value $\langle 
\mathcal{O} \rangle$ denotes $\textrm{Tr} [e^{-\beta\mathcal{H}_{\textrm{at}} }
S(\lambda)\mathcal{O}] /\mathcal{Z}_{\textrm{at}}$.  The column on the far 
right gives the results for the half-filled case considered here. The symbol 
$\chi_{\textrm{mixed}}$ denotes the mixed static local charge susceptibility
between the conduction and the localized electrons.
Recall at half filling $E_f=-U/2$, $w_1=1/2$, and $\rho_e=1/2$.}
\begin{ruledtabular}
\begin{tabular}{llll}
Moment&Operator Average&General result&Half-filling result\\
\colrule
$\int d\omega A_f(\omega)$&$\langle \{f,f^\dagger\}_+\rangle$&1&1\\
$\int d\omega A_f(\omega)f(\omega)$&$\langle f^\dagger f\rangle$&$w_1$&1/2\\
$\int d\omega A_f(\omega)\omega$&$-\langle [\mathcal{H},f]f^\dagger
\rangle+\langle [\mathcal{H},f^\dagger]f\rangle$&$E_f+U\rho_e$&0\\
$\int d\omega A_f(\omega)\omega f(\omega)$&$\langle [\mathcal{H},f^\dagger]f
\rangle$&$E_f w_1+U(\chi_{\textrm{mixed}}+w_1\rho_e)$&
$U\chi_{\textrm{mixed}}$\\
$\int d\omega A_f(\omega)\omega^2$&$\langle [\mathcal{H},[\mathcal{H},f]]
f^\dagger\rangle+\langle [\mathcal{H},[\mathcal{H},f^\dagger]]f\rangle$&
$E_f^2+(2E_f+U)U\rho_e$&$U^2/4$\\
$\int d\omega A_f(\omega)\omega^2f(\omega)$&$\langle [\mathcal{H},[\mathcal{H},
f^\dagger]]f\rangle$&$E_f^2w_1+(2E_f+U)U(\chi_{\textrm{mixed}}+w_1\rho_e)$&
$U^2/8$\\
\end{tabular}
\end{ruledtabular}
\end{table*}

We find that sometimes the $\delta$-extrapolation scheme does not further
improve the accuracy of the spectral function.  In that case, it is often
more accurate to use the result generated with the smallest 
$\Delta t_{\textrm{real}}$.  In other cases, we find that the exact
result for the lowest Matsubara frequency Green's function is bracketed by the
calculation with the smallest $\Delta t_{\textrm{real}}$ and the 
$\delta$-extrapolation result.  In that case, we can average those two
spectral functions in order to produce better agreement for $G_f(i\omega_0)$.
We call this extrapolation procedure Matsubara-extrapolation.  It sometimes
can improve the accuracy of the results.

As a general rule of thumb, if we can achieve accuracy of better than
1\% for all of the spectral moments, and we can achieve four digits of accuracy
for all of the Matsubara frequency Green's functions, then the resulting
DOS is numerically quite accurate.  The deviations from the exact result
are most likely occurring at small frequencies, where we need long-time data
to get an accurate Fourier transform, and at high frequencies, where the tails 
don't always decay exactly to zero.

\begin{figure}[htb]
\epsfxsize=3.0in
\epsffile{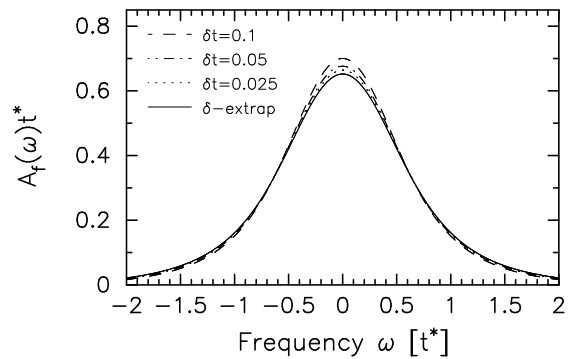}
\caption{\label{fig: u=1.5_T=5} 
$f$-electron DOS for different discretization sizes.  Also plotted is the
$\delta$-extrapolated result using a three-point (quadratic)
Lagrange interpolation formula. The parameters are $U=1.5$ on the Bethe lattice,
with $T=5.0$
}
\end{figure}

We illustrate these extrapolation procedures with the case $U=1.5$ on the Bethe
lattice (results for $U=1$ have also 
appeared\cite{freericks_turkowski_zlatic_2004b}).  
We first focus on high temperature, with $T=5$.  The results
for the moment sum rules, for the shift to the DOS, and for the lowest
Matsubara frequency Green's function are presented in 
Table~\ref{table: u=1.5_T=5}.  We plot the DOS for different 
$\Delta t_{\textrm{real}}$ in Fig.~\ref{fig: u=1.5_T=5}. As can be
seen in the figure, as the discretization size decreases, the DOS approaches
a limiting result, which is close to the one predicted by the
$\delta$-extrapolation procedure.  An examination of the table shows how
the moment sum rules and the Matsubara frequency Green's functions
are all improved as the discretization error is reduced.  The extrapolation
formula used a quadratic Lagrange interpolation with all the 
three DOS calculated
at different $\Delta t_{\textrm{real}}$'s. These results show that a systematic
extrapolation procedure is sometimes possible, and that the overall accuracy 
that can be achieved is quite high (of course it is difficult to estimate the
pointwise accuracy of the DOS from any of these integral sum rules).

\begin{table*}
\caption{\label{table: u=1.5_T=5}
Table of the accuracy of the different calculations of the DOS
by comparing results for the different sum rules ($U=1.5$ on the Bethe lattice,
with $T=5$).  The frequency cutoff for the
zero moment sum rule is $|\omega|<15$, while all other moments are cutoff at the
point where the integral stops increasing and approaches a constant (there
is usually a decrease for larger values of $\omega$) which normally
corresponds to $|\omega|<4$.}
\begin{ruledtabular}
\begin{tabular}{llllll}
Moment&$\Delta t_{\textrm{real}}=0.1$&$\Delta t_{\textrm{real}}=0.05$&
$\Delta t_{\textrm{real}}=0.025$&$\delta$-extrapolation&exact\\
\colrule
1&0.999994217&1.00000354&1.0000113&1.00000099&1\\
$f(\omega)$&0.502728929&0.501524381&0.500528293&0.49994769&0.5\\
$\omega f(\omega)$&-0.023149398&-0.025185717&-0.026169934&-0.02785382&-0.027912\\
$\omega^2$&0.465833709&0.507191087&0.527187957&0.56173037&0.5625\\
$\omega^2 f(\omega)$&0.232916854&0.253595543&0.263593979&0.28086519&0.28125\\
$G_f(i\omega_0)$&-0.063584452&-0.063571673&-0.063545431&-0.06351812&-0.063518334
\\
\colrule
shift&0.00485&0.005015&0.002534&-0.0000039&0\\
\end{tabular}
\end{ruledtabular}
\end{table*}

\begin{figure}[htb]
\epsfxsize=3.0in
\epsffile{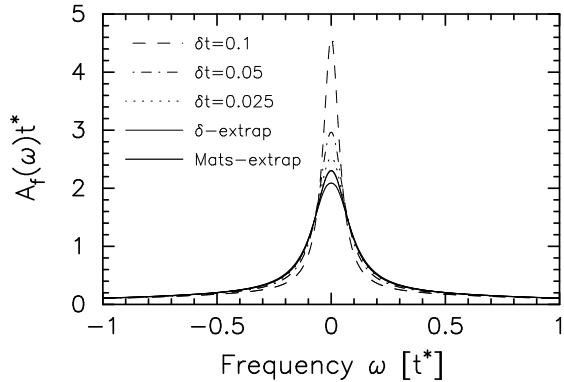}
\caption{\label{fig: u=1.5_T=0.1}
$f$-electron DOS for different discretization sizes.  Also plotted is the
$\delta$-extrapolated result using a linear interpolation formula for the
smallest two $\Delta t_{\textrm{real}}$ 
values and the Matsubara extrapolated 
result. The parameters are $U=1.5$ on the Bethe lattice, with $T=0.1$.
}
\end{figure}

As the temperature is lowered, the spectral function sharpens when $U$ is small.
In the noninteracting case, the spectral function is a delta function.  In the
interacting case, the spectral function approaches a delta function, but always
maintains a finite width, even at $T=0$. Nevertheless, the calculations become
more difficult at lower $T$, because a narrow peak in the DOS implies a slow
exponential decay in $G_f^>(t)$, and we find that the discretization
error also grows as $T$ is reduced.  To illustrate this phenomena, we 
show results for $U=1.5$ and $T=0.1$.  The moment sum rules are summarized
in Table~\ref{table: u=1.5_T=0.1} and the DOS are summarized in
Fig.~\ref{fig: u=1.5_T=0.1}.  One can see that as $\Delta t_{\textrm{real}}$
is made smaller, the peak in the DOS is reduced in height and increases in
width.  Furthermore, the $\delta$-extrapolation scheme seems to overcorrect,
by producing a DOS that is too wide (we use a two-point [linear] interpolation
formula here).  The Matsubara-extrapolation procedure
is much better, but the overall accuracy is reduced relative to the higher
temperature results (we find only about 3\% accuracy for the moments, and 
three parts in $10^3$ accuracy for the Matsubara frequency Green's
functions).  We find that this behavior is generic for our
calculations---usually the calculations are more difficult at lower temperature,
often requiring a smaller discretization size for the same level of
accuracy.  We also find that the real-frequency
extrapolation procedures start to break down as $T$ is reduced too.

\begin{table*}
\caption{\label{table: u=1.5_T=0.1}
Table of the accuracy of the different calculations of the DOS
by comparing results for the different sum rules ($U=1.5$ on the Bethe lattice,
with $T=0.1$).  The frequency cutoff for the
zero moment sum rule is $|\omega|<15$, while all other moments are cutoff at the
point where the integral stops increasing and approaches a constant (there
is usually a decrease for larger values of $\omega$) which normally
corresponds to $|\omega|<4$.}
\begin{ruledtabular}
\begin{tabular}{lllllll}
Moment&$\Delta t_{\textrm{real}}=0.1$&$\Delta t_{\textrm{real}}=0.05$&
$\Delta t_{\textrm{real}}=0.025$&$\delta$-extrapolation&Mats-extrapolation&
exact\\
\colrule
1&1.00015473&1.00000776&1.00000629&0.99999920&1.0000015&1\\
$f(\omega)$&0.503123093&0.501181165&0.500012329&0.50005666&0.5003576&0.5\\
$\omega f(\omega)$&-0.192872666&-0.208532743&-0.215836651&-0.22736437&-0.2227881
&-0.220742\\
$\omega^2$&0.477461453&0.518025518&0.537836969&0.58728661&0.5699565&0.5625\\
$\omega^2 f(\omega)$&0.238730727&0.259012759&0.268918484&0.29364330&0.2849782&
0.28125\\
$G_f(i\omega_0)$&-2.08188686&-1.96833103&-1.91591427&-1.86349693&
-1.8890850&-1.88908508\\
$G_f(i\omega_1)$&-0.882512945&-0.865141592&0.85643369&-0.847725223&-0.851976249
&-0.854845179\\
\colrule
shift&0.00501&0.005060&0.002544&-0.0000091&-0.0004000&0\\
\end{tabular}
\end{ruledtabular}
\end{table*}

\begin{figure}[htb]
\epsfxsize=3.0in
\epsffile{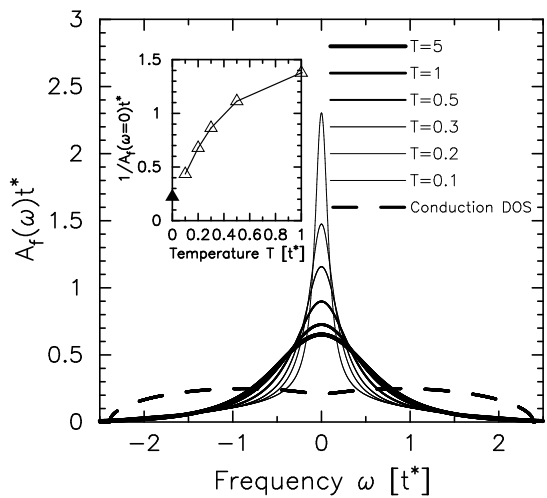}
\caption{\label{fig: u=1.5}
$f$-electron DOS for different temperatures.  Also plotted is the
conduction-electron DOS (which is temperature independent). 
The parameters are $U=1.5$ on the Bethe lattice. Inset is a plot
of $1/A_f(\omega=0)$ versus $T$.  Note how it appears to behave 
linearly at small $T$ allowing us to extrapolate to the $T=0$ result,
so we can predict the maximal height of the $f$-electron DOS at $T=0$.
}
\end{figure}

A summary of the results, for the case of $U=1.5$ on the Bethe lattice
is shown in Fig.~\ref{fig: u=1.5}.  We have used the most accurate DOS
calculated at each temperature, by one of the two extrapolation procedures.
We also included the conduction electron DOS, which has a dip develop
at the Fermi energy.  One can see that the $f$-electron DOS grows and
sharpens as $T$ is reduced.  We find that calculations at much lower 
temperatures than presented here become problematic due to discretization
and time-domain cutoff errors.  Inset into the figure is a plot
of $1/A_f(\omega=0)$ versus $T$.  We have linearly extrapolated the
last few points to estimate how big the DOS would grow as $T\rightarrow 0$.
Our estimate shows that the peak in the DOS should increase to about
$4.5$ as $T\rightarrow 0$.  Note the major differences between the
localized electron DOS and the conduction electron DOS.  The $f$-electron
DOS sharpens and concentrates much weight around $\omega=0$, while the
conduction electron DOS has a dip there.  Notice further that there is 
no significant change that we can see in our data near the band edge of the
conduction electron DOS that is also seen in the $f$-electron DOS, although
we expect that at $T=0$, the bandwidths of both DOS should agree with
each other.

\begin{figure}[htb]
\epsfxsize=3.0in
\epsffile{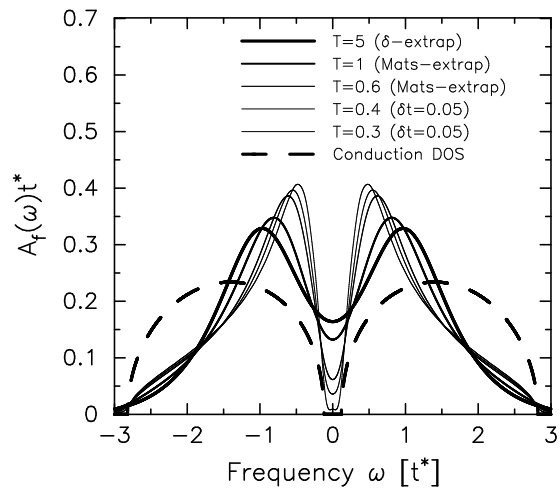}
\caption{\label{fig: u=2.5}
$f$-electron DOS for different temperatures with $U=2.5$ on the
Bethe lattice.  Also plotted is the
conduction-electron DOS (which is temperature independent).
The $T=5$ data uses $\delta$-extrapolation, the $T=1$ and 0.6 data use
the Matsubara-extrapolation procedure, and the lower temperatures
are not extrapolated (but have $\Delta t_{\textrm{real}}=0.05$).
Note how the $f$-electron DOS develops a gap as $T$ is lowered.  Note further
that a kink starts to develop near the upper and
lower band edges of the conduction DOS as expected too.  Our computational
accuracy is worst for the subgap DOS at low temperature.  
}
\end{figure}

We next investigate the case $U=2.5$ on the Bethe lattice (results for the
critical interaction strength $U=2$ appear 
elsewhere\cite{freericks_turkowski_zlatic_2004a}).  This case 
corresponds to lying just on the insulating side of the metal-insulator
transition (which occurs at $U=2$). The summary plot
of the DOS is presented in Fig.~\ref{fig: u=2.5}.  Note how the 
localized electron DOS sharpens and develops a gap as $T$ is lowered.
What is interesting, is that the DOS seems to pile up near the 
correlation-induced gap at low $T$.  We also see a kink start to develop
near the upper and lower conduction band edges, indicating that the $f$-electron
DOS will likely vanish outside of the band as $T\rightarrow 0$.  
Numerically, these calculations are challenging.  If the discretization error 
is too large, or the time-domain cutoff is too small, then we can find
negative DOS in the gap region at low temperature.  In fact, the poor quality of
our data for larger $\Delta t_{\textrm{real}}$ is the reason why we cannot
extrapolate the low temperature data faithfully. The accuracy of our
calculations is usually better than $1.5\%$ for the first moment, better than
$4\%$ for the second moment, and better than $0.5\%$ for the Matsubara frequency
Green's functions (at higher temperature, we do significantly better).

\begin{figure}[htb]
\epsfxsize=3.0in
\epsffile{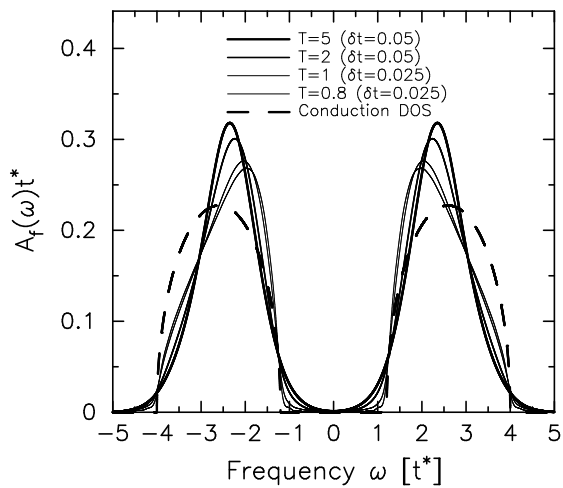}
\caption{\label{fig: u=5}
$f$-electron DOS for different temperatures with $U=5$ on the
Bethe lattice.  Also plotted is the
conduction-electron DOS. No extrapolations are used with this data,
just the lowest $\Delta t_{\textrm{real}}$ that produces a positive
DOS everywhere, and does not develop significant oscillations due
to the time-domain cutoff. Our computational resources don't allow
us to go below $T=0.8$ here.
}
\end{figure}

Next we consider a large-gap insulator on the Bethe lattice, with
$U=5$.  The summary plot is shown in Fig.~\ref{fig: u=5}.  Note how the
conduction-electron DOS has a large gap of about $2.5t^*$.  At high 
temperature, the $f$-electron DOS has significant subgap states.  As $T$ is
lowered, we find a transfer of spectral weight out of the gap region, with
the peaks moving towards the gap, and then some additional weight
being transfered to shoulders that lie close to the conduction band edges.
The small oscillations in the gap region for $T=0.8$ are artifacts of the 
cutoff in time.

We also present results for the hypercubic lattice.  The hypercubic lattice
does not develop gaps at the metal-insulator transition due to the infinite
exponential tails of the noninteracting Gaussian DOS.  But the spectral function
is suppressed to zero at the Fermi energy and there is a ``gap region'' where
the DOS remains exponentially small.  The transition occurs at $U=\sqrt{2}$,
and we expect results for the hypercubic lattice to be similar to those
of the Bethe lattice when $U_{\textrm{Bethe}}=\sqrt{2}U_{\textrm{hypercubic}}$
Brandt and Urbanek's original 
work\cite{brandt_urbanek_1992} presented results for the hypercubic lattice.  
Unfortunately they gave no details on the step sizes used in their computations
or of the accuracy of their results. The one discussion of moments that they
include gives an improper value to the second
moment of the $f$-electron DOS, and
it is likely they never checked the numerical accuracy of their results against
any moment sum rules.

We calculate three different values of $U$ for the hypercubic lattice:
$U=1$ which has a dip in the conduction electron DOS (similar to $U=1.5$
for the Bethe lattice), $U=2$ which is a ``small-gap'' insulator (similar to
$U=2.5$ on the Bethe lattice) and $U=4$ a ``large-gap'' insulator (similar
to $U=5$ on the Bethe lattice); the near-critical point $U=1.5$ appears
elsewhere\cite{freericks_turkowski_zlatic_2004a}. Brandt and Urbanek showed 
two DOS for $U=1$
and five DOS for $U=2$.  They did not calculate the $U=4$ case.

\begin{figure}[htb]
\epsfxsize=3.0in
\epsffile{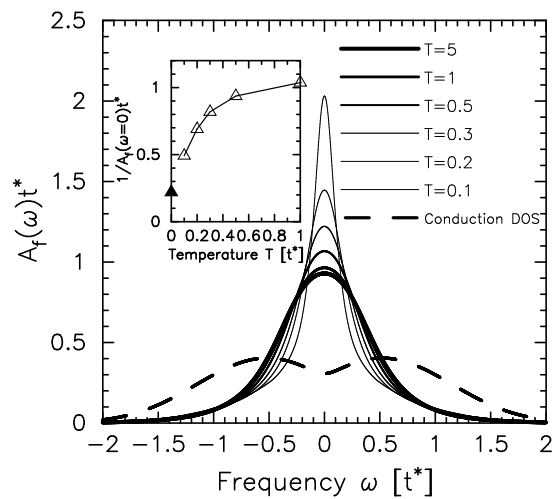}
\caption{\label{fig: u=1}
$f$-electron DOS for different temperatures with $U=1$ on the
hypercubic lattice.  Also plotted is the
conduction-electron DOS. All of the data was extrapolated with one of the
two extrapolation techniques discussed in the text. Note the similarity 
with Fig.~\ref{fig: u=1.5} for the Bethe lattice.  Inset we plot the
inverse of the DOS at the chemical potential.  Here the low-temperature
results don't appear to behave in quite the linear fashion we saw on the
Bethe lattice, but we can still attempt to extrapolate to $T=0$ with the
prediction that the peak in the DOS will also be around 4.5 at $T=0$.
}
\end{figure}

The $U=1$ case is plotted in Fig.~\ref{fig: u=1}.  The results shown here
are quite similar to those on the Bethe lattice (Fig.~\ref{fig: u=1.5}).
The DOS sharpens as $T$ is lowered, even though the conduction
electron DOS has a dip at the Fermi energy. Inset is a plot of the inverse of 
the DOS at the chemical potential versus $T$.  We can use it to extrapolate to 
$T=0$ and predict that the spectral function peaks at about 4.5.  Our results
at high temperature and at low temperature agree well with those
of Brandt and Urbanek\cite{brandt_urbanek_1992}.

\begin{figure}[htb]
\epsfxsize=3.0in
\epsffile{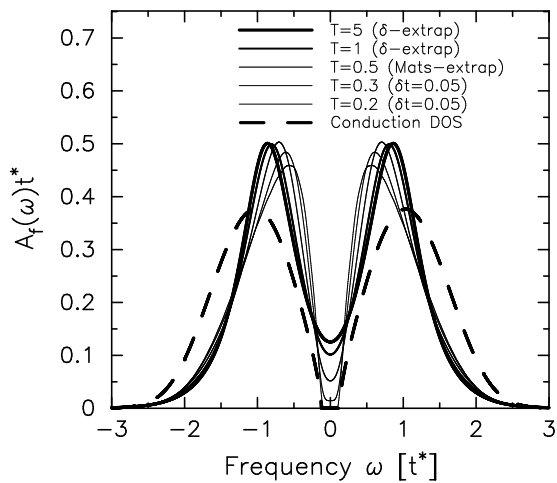}
\caption{\label{fig: u=2}
$f$-electron DOS for different temperatures with $U=2$ on the
hypercubic lattice.  Also plotted is the
conduction-electron DOS. The data is either extrapolated with one of the
two extrapolation techniques discussed in the text, or we work with a fixed
value of the discretization on the real-time axis. Note the similarity
with Fig.~\ref{fig: u=2.5} for the Bethe lattice.  
}
\end{figure}

Next we consider the case $U=2$ in Fig.~\ref{fig: u=2}, which should be
compared to the similar results on the Bethe lattice (Fig.~\ref{fig: u=2.5}).
Here we see the same kind of behavior---the gap is filled at high
temperature; as $T$ is lowered, spectral weight transfers from the 
gap region out to the band edges; and the peaks of the DOS migrate toward
the gap regions.  Note that the data shown for $T=0.2$ actually has a small
region of frequency where the DOS goes negative.  This is an artifact of
the discretization error and the time-domain cutoff.

\begin{figure}[htb]
\epsfxsize=3.0in
\epsffile{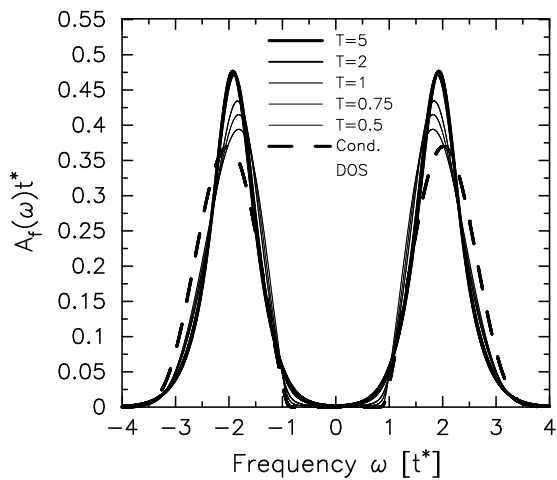}
\caption{\label{fig: u=4}
$f$-electron DOS for different temperatures with $U=4$ on the
hypercubic lattice (the $T=5$ and $T=2$ data overlap).  Also plotted is the
conduction-electron DOS.  The $T=5$ data is calculated with the
$\delta$-extrapolation technique; all other temperatures work with
$\Delta t_{\textrm{real}}=0.05$.
Note the similarity with Fig.~\ref{fig: u=5} for the Bethe lattice.
}
\end{figure}

The results for $U=4$ on the hypercubic lattice are presented in 
Fig.~\ref{fig: u=4}.  The behavior is what one expects: at high temperature,
the gap region is filled in by thermal excitations.  As the temperature is 
lowered, the gap region develops, with spectral weight being transferred from
the gap out to higher energy.  As the temperature becomes even lower, the 
computational needs exceed our resources.  Note how the peaks in the
$f$-electron DOS are pushed closer to the gap region, than the peaks in the
conduction-electron DOS.

\section{Conclusions}

In this contribution we calculated the $f$-electron DOS of the Falicov-Kimball
model at half filling.  The procedure requires us to generate the greater 
Green's function for real time and the Fourier transform to get the DOS.  Unlike
the conduction DOS, which is temperature independent, the $f$-electron DOS has
significant temperature dependence.  For small $U$, the DOS sharpens as $T$
is lowered to a single-peak structure with a narrow width.  For large $U$,
the DOS develops a gap at low $T$ and the peaks of the DOS push close to
the correlation-induced gap edges. When we compare results for similar $U$
values on the Bethe and hypercubic lattices, we see similar behavior in the DOS.

We performed an in depth analysis of the accuracy of the numerical calculations.
Errors arise from a finite discretization error (discretizing the continuous
matrix operator into a discrete matrix) and a time-domain cutoff error
[representing the largest time $t$ that $G_f^>(t)$ is calculated out to].
We use linear and quadratic moment sum rules and the spectral formula for
the Matsubara frequency Green's functions to gauge the accuracy of the 
calculations. In general, the discretization error becomes worse as
$T\rightarrow 0$ and it is quite challenging to get accurate results
at low temperatures and strong coupling.

This study is useful to understand problems with the accuracy of truly 
nonequilibrium calculations that use similar Kadanoff-Baym contours.  While we 
would not have exact sum rules to compare to anymore, it is clear that one 
needs to perform systematic studies in the discretization size along the
contour to gauge the accuracy of the results.  One also needs to reduce the 
real-axis discretization as the temperature is reduced.  
True nonequilibrium problems
evolve in an external field, and such a field can be added into the analysis
given here.  The complicated aspect is being able to construct the local
Green's function from the local self energy, as the coupling to a vector
potential enters into the hopping part of the Hamiltonian, and the local
Green's function is no longer represented by a simple Hilbert transform.

We only examined the half-filled case here.  This provides a significant 
simplification, as the DOS can be calculated by a Fourier transform of the
real part of $G_f^>(t)$.  For other fillings, the analysis is more complicated
and usually requires using particle-hole symmetry to generate the full DOS.
We plan to examine that case in the future.

\acknowledgments
We acknowledge support from the National
Science Foundation under grant number DMR-0210717 and the Office of Naval
Research under grant number N00014-99-1-0328.  
Supercomputer time was provided by the ERDC and ARSC supercomputer centers.

\bibliography{fk_dmft.bib}

\end{document}